\documentclass[preprint2,10pt,a4paper]{aastex}
\usepackage{amsmath}
\usepackage{graphicx}
\begin{document}
\title{Evolution of the bursting-layer wave during a Type 1 X-ray burst}
\author{R. G. Berkhout}
\affil{Leiden University, Leiden Observatory, P. O. Box 9513, NL-2300 RA Leiden}
\email{berkhout@strw.leidenuniv.nl}
\and
\author{Yuri Levin}
\affil{Leiden University, Leiden Observatory and Lorentz Instutute, P. O. Box 9513, NL-2300 RA Leiden}
\email{Levin@strw.leidenuniv.nl}
\begin{abstract} \noindent
In a popular scenario due
to Heyl, quasi periodic oscillations (QPOs) which are seen during  type 1 X-ray bursts are produced by giant travelling waves
in  neutron-star oceans. 
Piro and Bildsten have proposed that during the burst cooling the wave in the bursting layer may convert into a deep crustal interface wave, 
which would cut off the visible QPOs. This cut-off would help explain the magnitude of the QPO frequency drift, which is otherwise overpredicted
by a factor of several in Heyl's scenario.  
In this paper, we study the coupling between the bursting layer and the deep ocean. The coupling turns out to be weak and only 
a small fraction of the surface-wave energy gets transferred to that of the crustal-interface wave during the burst. Thus the crustal-interface
wave plays no dynamical role during the burst, and no early QPO cut-off should occur. 
\end{abstract}

\section{Introduction}
Unstable nuclear explosions of accreted hydrogen and helium on the surface of a neutron star (NS) 
produce  observable (type-1) X-ray bursts in some low-mass x-ray binaries.
The bursts typically have rise times of order $1~\hbox{s}$ and decay times of order $10-100~\hbox{s}$. 
During some of the bursts quasi-periodic oscillations (QPOs) are seen with frequencies ranging from $40$ to $700~\hbox{Hz}$. 
In the decaying tail of the burst the QPO frequency typically shows a positive drift of $1-5~\hbox{Hz}$ to an asymptotic value, which is 
characteristic to a bursting object \footnote{Notable exceptions are burst QPOs from millisecond X-ray pulsars. The QPO frequency of these 
objects stays constant during the bursts and is equal to the NS spin.}. 
This stability has led many to conclude that this asymptotic frequency is equal to the spin of the NS; in this picture the QPO is produced by
a rotating asymmetric temperature pattern on the NS surface. 
However,
after more than 10 years since the discovery of the burst QPOs,
it is still unclear what leads to the surface asymmetry at late times on these low magnetic-field NSs. 
The frequency drifts are also challenging to explain, since the NS spin should remain constant during the
burst. Strohmayer et al. 1997b argued that the conservation of angular momentum could be the cause of the drifts. 
In their scenario the spin of the bursting layer changes due to radial expansion and contraction during the burst. 
However detailed calculations show that this effect under-predicts the frequency drift in the burst tail  by a factor of 2-3 
(Cumming et al. 2002). It also does not explain the surface asymmetry at late times when the hot spot has already spread over the entire surface. 

In an important paper Heyl (2004) proposed the non-radial surface modes (waves) as the origin of the QPOs. 
In his scenario the burst excites  waves in the bursting layer, and it is these waves that modulate the X-ray flux periodically. 
The QPO's frequency drift can then be accounted for by the change of the wave frequency during the cooling of the bursting  layer.
 However, the basic surface wave frequencies change by order $10~\hbox{Hz}$ during the burst cooling, significantly larger than the observed amount. 
Two proposals have been put forward to overcome this difficulty. Firstly, Heyl (2004) has proposed that the modes localised in the photosphere will 
have smaller frequencies and drifts. However, due to the small density  and vertical scaleheight, the photospheric mode
frequencies are extremely sensitive to the presence of hydromagnetic stresses in the
neutron-star ocean. For example, even for a small vertical field of $B=10^5$G, the 
Alfven crossing frequency 
\begin{equation}
\nu_a\sim {B\over 2h\sqrt{4\pi\rho}}\sim 30 (B/10^5\hbox{G})(10\hbox{cm}/h)[10^4\hbox{g/cm}/\rho]^{1/2} \hbox{Hz},
\label{alfven}
\end{equation} 
is much greater than the frequency of the original mode. Here $\rho$ and $h$ are the characteristic density
and scaleheight of the photosphere. Thus even a small magnetic field is expected to strongly modify the photospheric
modes and substantially increase their frequencies.

Secondly, Piro and Bildsten (2005b) (hereafter P\&B05b) have suggested 
that the global surface wave is responsible for the QPOs, but that the frequency drift is cut short by resonant conversion of the 
bursting-layer wave (BLW) to the crustal-interface wave (CIW). The latter is an oscillatory mode whose energy is concentrated
 near the interface between the 
crystallised (crust) and liquid (deep ocean) burst ashes (Piro \& Bildsten 2005a [P\&B05a]). The CIW has a low frequency 
of $\sim 4$Hz, and its main restoring force is due to the shear modulus of the crystallised ashes. 
The frequency of the CIW is unperturbed during the burst, which implies a stable asymptotic frequency of the QPOs. 
One of the interesting features of this scenario is that the QPO asymptotic frequency differs from the NS rotational frequency by that of the CIW. 
This would have important implications for the narrow-band LIGO searches for the NS-generated gravitational waves.

In this paper we investigate in detail the P\&B05b scenario. 
First,
we evaluate the coupling between the bursting layer and the deep ocean in a shallow-water model. We find a good qualitative
agreement with modal P\&B05b calculations for a more realistic ocean model. Second,
we analyse in detail the dynamics of the
mode conversion.   P\&B05b have conjectured, without proof, that
the cooling period is long enough for the resonant wave conversion to occur. By contrast, we   
show that the cooling period is in fact {\it too short} for the resonant wave conversion to occur,  and thus that P\&B05b burst-QPO scenario is not viable. 


\section{The bursting-layer wave and the crustal-interface wave}
In Heyl's (2004) scenario a type 1 X-ray burst excites a backwards-propagating buoyant R-mode in the bursting layer. 
During the cooling period the pressure scale height of the bursting layer decreases. 
This makes the frequency of the wave decrease, which will be seen as a positive drift by an inertial observer. 
The deep ocean remains unperturbed during the burst so the frequency of the CIW is unchanged. 
At a certain point during burst cooling the wave frequencies will approach each other. 
However the two waves represent coupled vibrational modes of the NS. 
Because of the coupling, the vibrational mode frequencies do not cross but undergo an avoided crossing (P\&B05b). 
The mechanism for this phenomenon will be explained later on; 
for an illustration of an avoided crossing between modes see figure \ref{fig:avoidedc}. 
At the avoided crossing the two modes exchange their characteristics: the mode that is the BLW becomes the CIW and vice verse.
If the mode frequencies evolve sufficiently slowly (adiabatically) with time, the system, having started in one mode, remains continuously
in the same mode. This implies that upon a sufficiently slow passage through the avoided crossing, the BLW converts into CIW.  
P\&B05b have argued that this is what happens during the burst cooling, and they have argued that the conversion cuts short
the frequency drift of the BLW. While we agree with  P\&B05b about the location and strength of the avoided crossing, we show
below that the evolution is strongly non-adiabatic through the crossing, and hence that the BLW$\rightarrow$CIW conversion does not occur.

The BLW rides on top of the burst ashes. Taking a thin-ocean limit (see Pedlosky [1979]) 
\footnote{In this limit the radius of the NS is much bigger than the thickness of the bursting layer}, the dispersion relation is
\begin{equation} \label{eq:BLW}
\omega ^2_s \sim gH_bk^2 \frac{\Delta \rho}{\rho},
\end{equation}
where $g$ is the gravitational acceleration, $H_b$ is the pressure scale height at the interface
between the bursting layer and the ashes, $k$ is the effective wave number (to be defined later for the case
of rapid rotation), $\rho$ is the density of the ashes at the interface, and $\Delta\rho$ is the 
magnitude of the density jump across the interface.
 Here the $\Delta \rho/\rho$ 
 is important, since the density difference between the bottom of the bursting layer and top of the ocean becomes smaller during the burst 
cooling, which contributes to slowing down the wave. The radial component of the Coriolis force is
much smaller than the buoyancy force (Bildsten, Ushomirsky \& Cutler 1996, here after BUC96), 
which allows one to separate the angular and radial parts of the non-radial modes. Thus the mode solutions are as follows:
\begin{equation}
\eta(r,\theta,\phi,t)=g(r)k(\theta,\phi)\exp({-i\omega t}),
\end{equation}
where $r$ is the radius, $\theta$ is the latitude, $\phi$ is the longitude, $t$ is the time, $\eta(r,\theta,\phi,t)$ is the vibrational mode, 
$g(r)$ is the radial part of the mode, $k(\theta,\phi)$ is the angular part of the mode and $\omega$ is the angular frequency of the mode. 
Because the NS surface is azimuthally symmetric the azimuthal part is periodic:
\begin{equation}
\eta(r,\theta,\phi,t)=g(r)f(\theta)\exp{i(m \phi-\omega t)},
\end{equation}
where $f(\theta)$ is the latitudinal part of the mode and $m$ the azimuthal eigenvalue. The latitudinal part is given by the Laplace's tidal equation:
\begin{equation} \label{eq:tid}
\hat{L}_{q,m}\left[ f_{q,m}(\theta)\right] =-\lambda(q,m) f_{q,m}(\theta)
\end{equation} 
with $\hat{L}$ the Laplace tidal operator, $q=2\Omega/\omega$ the ratio 
between the NS rotation and the angular frequency of the mode, $\theta$ the latitude, 
$f$ a latitudinal eigenfunction and $\lambda$ the corresponding eigenvalue. The quantity $\lambda/R^2$ is the effective wavenumber $k$ for the waves
(Longuet-Higgins 1968). The mathematical expression for the Laplace tidal operator is as follows,
\begin{eqnarray}
\hat{L}_{q,m}=\frac{\partial}{\partial\zeta}\left( \frac{1-\zeta^2}{1-q^2\zeta^2}\frac{\partial}{\partial\zeta}\right) -\frac{m^2}{(1-\zeta^2)(1-q^2\zeta^2)}\\
-\frac{qm(1+q^2\zeta^2)}{(1-q^2\zeta^2)^2}, \nonumber
\end{eqnarray}
where $\zeta=\cos \theta$.
Longuet-Higgins (1968) found the angular eigenfunctions, 
called Hough functions, in terms of the spherical-harmonics series. Later BUC96 gave a simpler method to find the eigenfunctions numerically. 
To approximately match the QPO with the rotational frequency,
 the angular part should be that of the $m=1$ buoyant R-mode ($m=1$, $l=2$ R-mode for slow rotation; see Heyl 2004). 
This mode has $\lambda \sim 0.11$ in the fast rotating limit ($\omega \ll \Omega $; BUC96). 
P\&B05b estimated the initial frequency of the BLW to be around  $10.8~\hbox{Hz}$ in the frame  rotating with the NS,
 with a negative drift of about $9~\hbox{Hz}$ during the cooling period 
\footnote{To find these values they considered the bursting layer as an ideal gas, whose temperature is
set by a vertical radiative flux of $10^{25}~\hbox{ergs.cm}^{-2}.\hbox{s}^{-1}$. 
This layer extends till a depth where the density equals $10^6~\hbox{g/cm}^{3}$ and the pressure scale height is about $2~\hbox{m}$. 
During the burst cooling the temperature of the layer changes from order $10^9~\hbox{K}$ to order $10^8~\hbox{K}$. }.

The energy of the CIW is concentrated near the crustal interface with the deep ocean. The dispersion relation is 
\begin{equation}
\omega^2_c = \frac{\mu }{P} gH_ck^2,
\end{equation}
where $\mu$ is the shear modulus of the crust and $P$ and $H_c$ are the pressure and pressure scale height, respectively, at the crustal interface
(P\&B05a). The factor $\mu /P$ is present, since the crustal elasticity provides the restoring force for the CIW. 
Crystallisation of the electron degenerate ocean material is determined  by the dimensionless parameter 
\begin{equation}
\Gamma=\frac{\left( Z e \right)^2 }{a k_B T}
\end{equation}
where $k_b$ is Boltzmann's constant, $T$ is the temperature, $a=\left( 3/4 \pi n_i\right) ^2$, $n_i$ is the ion number density, $e$ 
is the electron charge and $Z$ is the average ion charge. Following Farouki \& Hamaguchi 1993, the crystallisation sets in when 
$\Gamma\simeq 173$, at a density of about $10^9~\hbox{g/cm}^{3} $. The pressure scale height at the crustal interface is of order $20~\hbox{m}$. 
P\&B05b have obtained  a frequency of $4.3~\hbox{Hz}$ for the 
CIW\footnote{The wave frequency was calculated for a NS accreting He-rich material at an accretion rate $\dot{M} \geq 10^{-9} \hbox{M}_ \odot$.}. 

\section{The shallow-water model}
P\&B05b have computed the mode frequencies of the coupled BLW and CIW and have demonstrated the existence of the avoided crossing.
In their computation, they have assumed a certain 1-dimensional model for the vertical density profile of the bursting layer.
However, this profile is poorly known in detail. The deflagration during the burst is a multi-dimensional phenomenon,
with most of the fuel ignited from the top by a strongly inclined deflagration front (Spitkovsky, Levin, \& Ushomirsky 2002),
and thus any 1-dimensional model is probably inaccurate. In this section we work out a simple  shallow-water
model for the coupling between layers, and find the strength of the
avoided crossing, and hence the coupling strength between the modes. Our results are in good qualitative 
agreement with P\&B05b; thus the existence and the strength of the
avoided crossing is a generic feature of the bursting ocean, and is insensitive to the uncertain details of the model.
The rest of this section is rather technical, and can be safely skipped by the reader not interested in the
mathematical details.

\begin{figure} [h]
\includegraphics[scale=0.4]{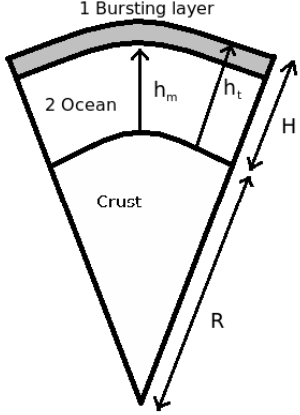} 
\centering
\caption{\textit{This image shows a cut out of the shallow-water model. 
Two layers on a large spinning star representing the bursting layer and the deep ocean of NS surface. 
The bursting layer has height $h_t$ and the deep ocean has height $h_m$ above the crust.}}
\label{fig:mod}
\end{figure}

Our basic idea is to represent the bursting layer and deep ocean by two layers of incompressible inviscid fluid (figure \ref{fig:mod}). 
The bottom beneath the two layers is flexible, and  represents the crust.
The top of the bursting layer is at the radius $R+h_t$, the top of the deep ocean is at $R+h_m$ and the crustal interface is at $R+h_b$ ($h_b=0$ in equilibrium state). For this model we make use of the shallow-ocean approximation. In this situation the thickness of the fluid layers, which is of order $H$, is small compared to the radius $R$ of the NS, i.e.~we assume $\delta=H/R\ll1$. 
The dynamical equations of motion for the thin water layers are the continuity equation for an incompressible fluid
\begin{equation} \label{eq:1ss}
\nabla \cdot \vec{v}=0
\end{equation}
and the momentum equation for an inviscid fluid
\begin{equation} \label{eq:2ss}
\frac{d\vec{v}}{dt}+2\vec{\Omega}\times \vec{v}=\frac{-\nabla P}{\rho}-\nabla \Phi,
\end{equation}
where $\vec{v}$ is the velocity and $\Phi$ is the gravitational potential. There is no exchange of fluid between the layers.
 Following Pedlosky (1979) and expanding Eq.~(\ref{eq:2ss}) to zeroth order in $\delta$ we find the radius-independent 
equations for the transverse 
velocity:
\begin{subequations}
\begin{align}
&\frac{dv_{\theta 1}}{dt}+2\Omega v_{\phi 1} \cos \theta = \frac{-g\partial_\theta h_t}{R}, \label{eq:3ss} 
\\
&\frac{dv_{\theta 2}}{dt}+2 \Omega v_{\phi 2} \cos \theta =\frac{-g\partial_\theta h_m}{R}-\frac{g \rho_1}
{R \rho_2} \partial_\theta (h_t - h_m),  \label{eq:4ss}
\\
&\frac{dv_{\phi 1}}{dt}-2\Omega v_{\theta 1} \cos \theta = \frac{-g \partial_\phi h_t}{R \sin \theta},  \label{eq:5ss}
\\
&\frac{dv_{\phi 2}}{dt}-2 \Omega v_{\theta 2} \cos \theta =\frac{-g\partial_\phi h_m}{R \sin \theta}-\frac{g \rho_1 \partial_\phi (h_t - h_m)  }
{R \rho_2 \sin \theta}. \label{eq:6ss}
\end{align}
\end{subequations}
Here the labels $1$ and $2$ stand for the top and the bottom layers, respectively.
To solve for the radial velocity we substitute Eqs.~(\ref{eq:3ss}), (\ref{eq:4ss}), (\ref{eq:5ss}) and (\ref{eq:6ss}) 
in the continuity equation (\ref{eq:1ss}) and the boundary conditions mentioned above. We keep the  terms up to the
 first order in $\delta$, since the radial velocity is of this  order relative to the horizontal one. We find
\begin{subequations}
\begin{eqnarray} \label{eq:vr1}
&&v_{r1}=\frac{B_1}{R}\left[ h_m-y \right] +\frac{dh_m}{dt},
\\ \label{eq:vr2}
&&v_{r2}=\frac{B_2}{R}\left[ h_b-y \right] +\frac{dh_b}{dt},
\end{eqnarray}
\end{subequations}
where $y$ is the height above the bottom of the shallow-water model, so that $r=R+y$. From the continuity equation, we find
\begin{subequations}
\begin{align} \label{eq:mc1}
\frac{dh_t}{dt} - \frac{dh_m}{dt}=\frac{B_1}{R}\left[  h_m - h_t  \right], 
\\ \label{eq:mc2}
\frac{dh_m}{dt} - \frac{dh_b}{dt}=\frac{B_2}{R}\left[ h_b -h_m  \right], 
\end{align}
\end{subequations}
where $B$ is the angular part of the velocity divergence
\begin{equation} \label{eq:B}
B= \left[ \frac{\partial_\phi v_{\phi }}{\sin \theta}+\partial_\theta v_{\theta }+\cot \theta v_{\theta }\right].
\end{equation}
Now we work out the modes using a linear perturbation theory. We look for solution in the following form:
\begin{equation} \label{eq:per}
Q(r,\theta,\phi,t)=Q_0(r)+Q' (r,\theta)\exp{i[m \phi-\omega t]},
\end{equation} 
where $Q$ is a dynamical variable, $Q_0$ is its
 equilibrium value,  and $Q'$ is the Eulerian perturbation amplitude. 
 Using Eq.~(\ref{eq:per}) in the Eqs.~(\ref{eq:3ss}) and (\ref{eq:5ss}) and keeping only linear  terms we find
\begin{subequations}
\begin{align} \label{eq:7ss}
&\omega^2\xi_{\theta 1}-i 2\omega \Omega \cos \theta \xi_{\phi 1} =\frac{-g \partial_\theta h'_t}{R},
\\ \label{eq:8ss}
&\omega^2\xi_{\phi 1}+i 2\omega \Omega \cos \theta \xi_{\theta 1} =\frac{-ig m h'_t}{R\sin \theta},
\end{align}
\end{subequations}
where $\xi$ is the amplitude of the Lagrangian
 displacement vector. Using Eqs.~(\ref{eq:7ss}) and (\ref{eq:8ss}) we can solve for $\xi_{\phi 1}$ and $\xi_{\theta 1}$ in
 terms of $h'_t$. Perturbing equation (\ref{eq:B}) and expressing $\xi_{\phi 1}$ and $\xi_{\theta 1}$ in terms of $h'_t$, we get
\begin{equation} \label{eq:B1}
B'_1=\frac{-ig}{R\omega} \hat{L}\left[h'_t \right]. 
\end{equation}
Analogous to the derivation for layer 1 we find for layer 2
\begin{equation} \label{eq:B2}
B'_2=\frac{-ig}{R\omega} \hat{L}\left[h'_m+\frac{\rho_1}{\rho_2}(h'_t-h'_m) \right]. 
\end{equation} 
From Eqs.~(\ref{eq:mc1}) and (\ref{eq:mc2}) we find 
\begin{subequations}
\begin{align}\label{eq:9ss}
&i\omega \left[ h'_m-h'_t\right] =\frac{B'_1}{R}\left[ h_{m0}-h_{t0}\right], &
\\  \label{eq:10s}
&i\omega \left[ h'_b-h'_m\right] =\frac{B'_2}{R}\left[h_{b0}-h_{m0}\right]. &
\end{align}
\end{subequations}
These differential equations can be separated in a radial part and an angular part. 
The angular part is Laplace's tidal equation (see eq.~[\ref{eq:tid}]). 
Rewriting Eqs.~(\ref{eq:9ss}) and (\ref{eq:10s}) using Eqs.~(\ref{eq:B1}), (\ref{eq:B2}) and (\ref{eq:tid}) 
we find the frequency of the modes. There are two equations
\begin{subequations}
\begin{align}  \label{eq:fr1}
\omega^2=\frac{g\lambda(q,m)\left[ h_{t0}-h_{m0}\right] }{R^2}\frac{1}{\left( 1-\frac{h'_m}{h'_t}\right)} ,\hbox{ and}
\\  \label{eq:fr2}
\omega^2=\frac{g \lambda(q,m) \left[ h_{m0}-h_{b0}\right] }{R^2}\frac{ \left[  \frac{h'_m}{h'_t}
 +\frac{\rho_1}{\rho_2}\left( 1-\frac{h'_m}{h'_t}\right)  \right]}{\left( \frac{h'_m}{h'_t}-\frac{h'_b}{h'_t}\right) }.
\end{align}
\end{subequations}
 In the shallow-water model the depth within a layer is equal to the pressure scale height. 
Thus, as the term with  $(h_{t0}-h_{m0})$ indicates, Eq.~(\ref{eq:fr1}) describes a wave concentrated 
in the bursting layer, i.e. the BLW. The vertical displacement of the top of the ashes affects 
the frequency of this wave,  as is represented 
 in Eq.~(\ref{eq:fr1}) by the term $h'_m/h'_t$. Eq.~(\ref{eq:fr2}) describes a wave 
 concentrated in the deep ocean, i.e.~the CIW. 
The flexing of the crust is represented by the term $h'_b/h'_t$ in Eq.~(\ref{eq:fr2}).

\subsection{Finding the avoided crossing}
During a type 1 X-ray burst the bursting layer expands and a density discontinuity is created at the bursting layer and ocean interface. 
This density discontinuity produces a buoyancy force at this interface, which acts as the restoring force for the BLW. 
The dispersion relation for the BLW is given by 
Eq.~(\ref{eq:BLW}). The equation contains a factor $\Delta\rho/\rho$,
 which mimics the strength of the buoyancy at the top of the ashes. 
The standard dispersion relation for a shallow-water wave is
\begin{equation}
\omega^2=gHk^2,
\end{equation}
which applies to waves in the shallow-water model. This equation does not contain the factor $\Delta\rho/\rho$ of Eq.~(\ref{eq:BLW}). 
To let the model take this term into account we define the effective pressure scale height of the BLW to be:
\begin{equation}
H_{\rm eff}=H_b\frac{\Delta\rho}{\rho}.
\end{equation}
In the shallow-water 
model we set the pressure 
scale height of the bursting layer to $H_{\rm eff}$, rather than $H_b$. During the burst cooling on the NS the pressure scale height of the 
bursting layer and the density discontinuity at the bursting layer and ocean interface decrease. 
To imitate both these effects in the model we simply decrease the value of the pressure scale height of the top layer.

We use the following representative parameters for our shallow-water model.
\begin{enumerate}
\item The angular frequency of the NS is $\Omega \sim 1000~\hbox{rad.s}^{-1}$.
\item On the surface of the NS the centripetal force is just one percent of the strength of the gravitational force. We approximate the radial acceleration as the gravitational acceleration; $g \sim 1.858\times10^{14} ~\hbox{cm.s}^{-2}$.
\item The mass of the bursting layer is $2.55\times10^{21}~\hbox{g}$.
\item The mass of the deep ocean is $2.52\times10^{25}~\hbox{g}$.
\item The pressure scale height of the bursting layer is set to 2.28m to give the BLW a frequency of $10.8$ Hz.
\item The pressure scale height of the deep ocean is set to 40.59 m to give the CIW a frequency of $4.3$ Hz.
\item The angular eigenvalue $\lambda(q,m)$ is set to that of the m=1 buoyant R-mode.
\end{enumerate}
For the calculations of the masses of the layers, we used an equation of state with $P \sim \rho^{{4}/{3}}$. This applies to the burning layer, 
which consists of an ideal gas and where the flux is approximately constant within the layer, and to 
the deep ocean, which has a pressure dominated by relativistic degenerate electrons. The equation of hydrostatic equilibrium is
\begin{equation}
\frac{dP}{dr}=-g\rho
\end{equation}
With this equation and the equation of state of above the density can be rewritten as a function of the radius
\begin{equation}
\rho\sim a(R-r)^3,
\end{equation}
where $a$ is some constant. We solved for $a$ using the NS surface parameters given by P\&B05b,
 and by integrating over the density we found the masses of the layers.
To find the latitudinal eigenvalue $\lambda$ for the $m=1$ buoyant R-mode, we solved 'Laplace Tidal equation' using the numerical
method described in BUC96.

A pressure perturbation in the deep ocean causes the neutron star crust to flex. To incorporate this flexing into our model we followed the analysis of P\&B05a. 
We assumed that the ocean is plane-parallel instead of spherical and we denote the horizontal and vertical coordinates by x and z, respectively. 
The bottom of the model represents the crustal interface of the NS surface. The $zz$ component of the Lagrangian perturbation of the stress tensor $\Delta \sigma_{zz}$, must be continuous at the crustal interface to avoid infinite vertical acceleration. On the crustal side of the interface, we have
\begin{equation} \label{eq:bel}
\Delta \sigma_{zz}^{below} = \frac{3\Gamma_1 /4+\mu_0}{\left( 1-\frac{h_{t0}}{s_{end}}\right)^3 }h'_bg\rho_2,
\end{equation}
where $\Gamma_1$ is the adiabatic exponent and $s_{end}$ is the depth from the surface where the CIW ends ($\vec{\epsilon}(s_{end})=0$). On the fluid side of the crustal interface,
\begin{equation} \label{eq:abo}
\Delta \sigma_{zz}^{above} = -\rho_1 g (h'_t-h'_m)-\rho_2 g (h'_m-h'_b).
\end{equation}
Equalising equations (\ref{eq:bel}) and (\ref{eq:abo}), setting $\Gamma \sim 4/3$ and taking $s_{end}>>h_{t0}$ (see P\&B05a) we find the following relation for the bottom flexing
\begin{equation} \label{eq:mub}
\mu_0\frac{h'_b}{h'_t}=-\frac{\rho_1}{\rho_2}\left( 1-\frac{h'_m}{h'_t}\right) -\frac{h'_m}{h'_t}.
\end{equation}
We express ${h'_b}/{h'_t}$ in terms of ${h'_m}/{h'_t}$ using Eq.~(\ref{eq:mub}) and solve Eqs.~(\ref{eq:fr1}) and (\ref{eq:fr2})
 for the frequencies of the modes and ${h'_m}/{h'_t}$.

\begin{figure}[h]
\includegraphics[scale=0.65,angle=270]{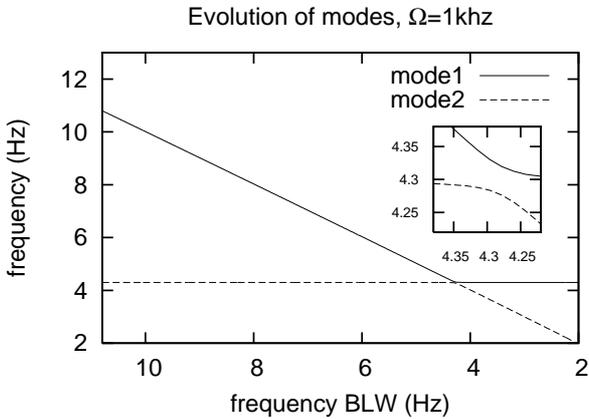}
\centering
\caption{\textit{In this figure the frequencies of the 2 
modes of the model have been plotted versus the frequency of the 
uncoupled BLW mode, $\bar{\omega}_1/(2\pi)$.
 The modes undergo an avoided crossing. At the avoided crossing mode 1 changes from the BLW to the CIW and mode 2 vice verse.}}
\label{fig:avoidedc}
\end{figure}
In figure \ref{fig:avoidedc} the  frequencies of the two modes have been plotted versus
the , uncoupled BLW mode, $\bar{\omega}_1/(2\pi)$, for a NS spinning at $1000~\hbox{rad/s}$. 
We determine the presence of an  avoided crossing, and measure the minimal frequency difference (i.e., the gap) between the modes to be $0.042$Hz. This
is similar, in order of magnitude, to that in P\&B05b, who have measured $0.058$Hz for a more realistic model of the NS ocean. 

In the next chapter we will use the size in the gap of the avoided crossing to find the coupling between the bursting layer and deep ocean,
and to model the dynamics of the system as it passes through the avoided crossing. 

\section{Dynamics of transition through the avoided crossing}
The avoided crossing found in the previous section results from the coupling between the top and the bottom layers. Without the coupling, each 
layer would have a vibrational mode frequency set by the individual dispersion relation for shallow-water waves. 
When coupling is introduced the layers push and pull each other. However, away from the avoided crossing, the coupling is weak and the modes are 
identified with the individual layers. P\&B05b argued that as $H_{\rm eff}$ is slowly reduced, at the avoided crossing the two modes would adiabatically 
exchange their characteristics. 
The mode concentrated in the bursting layer would transfer its energy to the deep ocean and vice verse. 
P\&B05b have used the following criterion for the adiabaticity: $\tau\gg 1/\omega$, where 
$\tau$ is the characteristic timescale of the frequency change. This criterion is fine {\it away}
from the crossing; however, as is well-known from quantum mechanics, it {\it breaks down} near
the avoided crossing. The theory of passing through the avoided crossing
was worked out independently by Landau (1932) and Zener (1932). In this section we first
model explicitly the mode crossing numerically, and then show that our results
are consistent with the Landau-Zener theory. Our conclusions are opposite to  those reached by
P\&B05b: the burst cooling is not slow enough for the modes to adiabatically exchange their characteristics. 

Let $x_1$ and $x_2$ be the generalised vibrational amplitudes of the two modes localised in the bursting layer and deep ocean respectively. 
These modes are coupled via a general interaction Hamiltonian $H_{\rm int}=-\alpha x_1 x_2$, and their joint equations of motion are
\begin{subequations} 
\begin{align}
&m_1\ddot{x}_1+k_1x_1=\alpha x_2, 
\\
&m_2\ddot{x}_2+k_2x_2=\alpha x_1,
\end{align}
\end{subequations}
where $m_1$ is the effective mass of the mode in the bursting layer, $m_2$ is the effective mass of the mode in the deep ocean, ${k_i}/{m_i}$ 
is the vibrational frequency of the layers set by the dispersion relation for shallow-water waves. It is convenient to rewrite these equations
as follows:
\begin{subequations} 
\begin{align} \label{eq:osc1}
&\ddot{\bar{x}}_1+\bar{\omega}_1^2\bar{x}_1=\bar{\alpha} \bar{x}_2,
\\ \label{eq:osc2}
&\ddot{\bar{x}}_2+\bar{\omega}_2^2\bar{x}_2=\bar{\alpha} \bar{x}_1,
\end{align}
\end{subequations}
where $\bar{x}_i=x_i\sqrt{m_i}$, $\bar{\alpha}={\alpha}/{\sqrt{m_1m_2}}$ and $\bar{\omega}_i={k_i}/{m_i}$. The two proper frequencies of the system are given by
\begin{equation}
\omega_\pm=\sqrt{\frac{\bar{\omega}_1^2+\bar{\omega}_2^2}{2} \pm \sqrt{\bar{\alpha} ^2+(\bar{\omega}_1^2-\bar{\omega}_2^2)^2}}.
\end{equation}
Note that
\begin{equation}
\frac{1}{2}\left( \omega^2_+ - \omega^2_-\right) = \sqrt{\bar{\alpha} ^2+(\bar{\omega}_1^2-\bar{\omega}_2^2)^2}.
\end{equation}
Therefore, the effective coupling $\bar{\alpha}$ can be expressed as
\begin{equation}
\frac{\bar{\alpha}}{\bar{\omega}_1^2}=\min\left[\frac{1}{2} \frac{\left( \omega^2_+ - \omega^2_- \right)}{\bar{\omega}_1^2}\right] 
\end{equation}
Thus $\frac{\bar{\alpha}}{\bar{\omega}_1^2}$ can be read off directly from the plot in Fig.~\ref{fig:avoidedc}:
\begin{equation} \label{eq:min}
\frac{\bar{\alpha}}{\bar{\omega}_1^2}\sim \frac{\Delta \omega}{\omega},
\end{equation}
where $\Delta \omega$ is the minimal separation between the two branches. Using Eq.~(\ref{eq:min}) and the data of 
Fig.~\ref{fig:avoidedc} we find the coupling between the layers for different rotational spin of the NS which is shown in table \ref{table1}.  
\begin{table}[h]
\center
\begin{tabular}{|c|c|}
\hline
Spin (rad.s$^{-1}$) $(\Omega)$ & coupling ($\hbox{rad.s}^{-2}$) $\left(\bar{\alpha}\right)$  \\
\hline
1000 & 7.29 \\ \hline
4000 & 7.41\\
\hline
\end{tabular}
\caption{\textit{This table shows the effective coupling coefficient between the bursting layer and deep ocean.}}
\label{table1}
\end{table}
Compared to the angular frequencies of the BLW and CIW the coupling is rather small, i.e. $\frac{\bar{\alpha}}{\bar{\omega}_1^2} \ll 1$. 
This is because of the huge effective mass difference between the two layers: 
the deep ocean is $\sim 10^4$ times heavier than the bursting layer.

We now integrate Eqs.~(\ref{eq:osc1}) and (\ref{eq:osc2}) numerically, using the values of $\bar{\alpha}$ in table \ref{table1}. We 
set the initial frequencies of oscillator 1 and 2 equal to the frequencies of the bursting layer wave ($10.8~\hbox{Hz}$) and the crustal interface 
wave ($4.3~\hbox{Hz}$), respectively. To simulate the burst cooling we let the frequency of oscillator 1 decrease 
exponentially by $9~\hbox{Hz}$ during a time period set by the typically observed cooling period of $10~\hbox{s}$. More precisely, we
evolve the frequency of oscillator 1 as a function of time as follows: first, we keep it constant
for 10 seconds, then we evolve it for 10 seconds according to the following expression
\begin{equation}
\frac{\bar{\omega}_1(t)}{2\pi}=14.24 \hbox{ Hz}\exp\left[ {\frac{-(t-10\hbox{s})}{10\hbox{ s}}}\right] -3.44\hbox{ Hz},
\end{equation}
and then we keep it constant for another 10 seconds.
\begin{figure} [h]
\includegraphics[scale=0.65,angle=270]{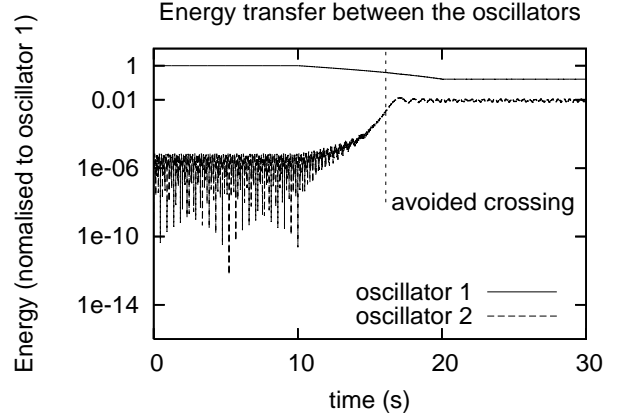}  
\centering
\caption{\textit{This figure shows the energy transfer between two coupled harmonic oscillators. During the first and last 10 s the frequencies of the oscillators are constant, but in the range $10-20~\hbox{s}$ the frequency of oscillator 1 goes down crossing the frequency of oscillator 2.}}
\label{pic:osc}
\end{figure}
In figure \ref{pic:osc} we plot the energy of the two oscillators normalised to the initial energy of oscillator 1, over the whole of the $30~\hbox{s}$ 
time interval. 
Oscillator 1 starts with a certain amplitude, while oscillator 2 begins with zero displacement and velocity. 
Because initially the coupling between the oscillators is small both oscillators are close 
to vibrational eigenmodes of the system, with oscillator 1 representing the excited BLW at the start of the burst cooling period. 
During the first $10~\hbox{s}$ virtually no energy is transferred between the oscillators. 
In the second interval of $10~\hbox{s}$ the energy of oscillator 1 decreases and the energy of oscillator 2 increases,
 especially at the location of the avoided crossing. However the energy of oscillator 2 remains much smaller than the energy of oscillator 1. 
In the last $10~\hbox{s}$ again there is virtually no energy transfer. 
We conclude that during the burst cooling the system does not evolve adiabatically and that the BLW does not change into a CIW.

We have run the simulation for several different values of $\mu_0$, $\Omega$ and ${m_1}/{m_2}$, and in each case found that the 
energy transfer between the oscillators is very small.
 The most important parameter is the cooling time; with cooling times $\geq 100\hbox{ s}$ we get a significant decrease for the BLW energy. 
However, this is far longer than most observed cooling times. Thus mode conversion cannot be a robust feature of the  Type 1 X-ray bursts.

We now compare our results with the Landau-Zener theory (Landau 1932, Zener 1932; for a quick derivation,
see Wittig 2005). 
In the Landau-Zener problem the Hamiltonian is changing so that the difference between the energy levels of a quantum-mechanical
system  is decreasing linearly with time and then crossing zero. 
This gives the system, which is originally in the first energy state, a chance to make a transition to the second energy state. 
The expression for the transition probability is 
\begin{equation}
p_{1\rightarrow2}=\exp{\left[ -\frac{\pi}{2} {\Delta \omega^2}\left({\frac{d(\bar{\omega}_1-\bar{\omega}_2)}{dt}}\right)^{-1}\right]} ,
\end{equation}
Using the formula we find that $p_{1\rightarrow2}$ is close to unity for our parameters, and thus that resonant 
conversion is not expected to occur. The conversion would occur if $p_{1\rightarrow2}$ were
to remain close to zero; for this the cooling time greater than $\sim 100$s is required,
in agreement with our numerical calculations.

\section{Outlook}
In this paper we have shown that the bursting-layer wave does not turn into the crustal-interface wave during Type 1 X-ray burst cooling. 
This argues against the P\&B05b picture in which the asymptotic frequency of the burst QPO differs from the NS spin by the crustal-interface 
wave frequency ($4~\hbox{Hz}$). Our conclusions are consistent with a very recent discovery  that
the accreting neutron star Aql X-1 is a transient millisecond x-ray pulsar (Casella et al., 2007). In that object, the asyptotic maximal frequency of
the burst QPOs is very close ($\sim 0.5$Hz) to the  spin frequency of the neutron star, without any shift due to the CIW.
We draw two  astrophysical conclusions from our calculations. Firstly, the 
simple surface waves do not work as the source of the burst QPO's. To find the source one should appeal to the waves in the photosphere (Heyl 2004) or 
some other non-trivial modes. Secondly,  the asymptotic spin of the QPOs is likely to be close to the NS spin, without
modification due to the CIW motion. 
Advanced LIGO should take this into account for the future searches of gravitational waves from accreting and bursting NSs.
\section{Aknowledgements}
After this paper was completed, we have learned that Tony Piro and Lars Bildsten have independently obtained similar results.
We thank Tony Piro for helpful discussions.
\section{References}
\begin{footnotesize} \noindent
Bildsten, L., Ushomirsky, G., \& Cutler, C. 1996, ApJ, 460, 827 (BUC96)\\
Brekhovskikh, L. M., \& Goncharov, V. 1994, Mechanics of Continua and Wave Dynamics (Berlin:Springer)\\
Casella, P., et al.~2007, submitted to ApJ Letters, arXiv:0708.1110 (astro-ph)\\
Chapman, S., \& Lindzen, R. S. 1970, Atmospheric Tides (Dordrecht:Reidel)\\
Cumming, A. et al. 2002, ApJ, 564, 343\\
Farouki, R., \& Hamaguchi, S. 1993, Phys. Rev. E, 47, 4330\\
Heyl, J.S. 2004, ApJ, 600, 939\\
Landau, L.D.~1932, Phys.~Z., 2, 46\\
Lee, U. 2004, ApJ, 600, 914\\
Longuet-Higgins, M. S. 1968, Phil. Trans. R. Soc. London, 262, 511\\
Pedlosky, J. 1979, Geophysical Fluid Dynamics (New york:Springer)\\
Piro, A. L. , \& Bildsten, L. 2005a, ApJ, 619, 1054 (P\&B05a)\\
Piro, A. L. , \& Bildsten, L. 2005b, ApJ, 629, 438 (P\&B05b)\\
Spitkovsky, A., Levin, Y., \& Ushomirsky, G. 2002, ApJ, 566, 101\\
Strohmayer, T. E. et al. 1991, ApJ, 375, 679\\
Strohmayer, T. E. et al. 1997b, ApJ, 486, 355\\
Strohmayer, T. E. et al. 1997a, ApJ, 487, L77\\
Wittig, C.~2005, J.~Phys.~Chem.~B, 109, 8428\\
Zener, C.~1932, Proc.~R.~Soc.~London A, 137, 696\\
\end{footnotesize}
\end{document}